\documentclass[letterpaper, 10 pt, conference]{ieeeconf}  %

\IEEEoverridecommandlockouts                              %

\usepackage{graphicx}      %
\usepackage[table,xcdraw]{xcolor}
\usepackage{amsmath}       %
\usepackage{amsfonts}      %
\usepackage{algorithm}     %
\usepackage{todonotes}
\usepackage{placeins}
\usepackage{siunitx}
\usepackage{textpos}

\newcommand{\argmin}[1]{\mathrm{arg}\min_{#1}} %

\newcommand{\params}{\boldsymbol{\theta}} %
\newcommand{\objSet}{\mathcal{J}}
\newcommand{\paramSet}{\Theta}
\newcommand{\T}{\mathbf{T}}
\newcommand{\g}{\mathbf{g}}
\newcommand{\lam}{\boldsymbol \lambda}
\newcommand{\mdot}{\dot m}

\newcommand{\cnstrntSet}{\mathcal{G}}
\newcommand{\contextSet}{\mathcal{S}}
\newcommand{\context}{\mathbf{s}}
\newcommand{\norm}[2]{\left\lVert#1\right\rVert_{#2}}
\title{\LARGE \bf
Vehicle Cabin Climate MPC Parameter Tuning Using Constrained Contextual Bayesian Optimization (C-CMES)
}

\author{David Stenger$^{1,\ast} $, Tim Reuscher$^{1,\ast}$, Heike Vallery$^{1,2}$, and Dirk Abel$^{1}$%
\thanks{$^{1}$ David Stenger, Tim Reuscher, Heike Vallery, and Dirk Abel are with the Institute of Automatic Control, RWTH Aachen University, {\tt\small \{d.stenger, t.reuscher\}@irt.rwth-aachen.de}}
\thanks{$^2$ Heike Vallery is also with the Department of BioMechanical Engineering, Delft University of Technology, and with the Department for Rehabilitation Medicine, Erasmus MC, Rotterdam, The Netherlands.}
\thanks{$^{\ast}$ These authors contributed equally and are listed in random order.}%
}

\begin{document}

\maketitle
\thispagestyle{empty}
\pagestyle{empty}

\begin{abstract}

       Climate-controlled cabins have for decades been standard in vehicles. Model Predictive Controllers (MPCs) have shown promising results in achieving temperature tracking in vehicle cabins and may improve upon model-free control performance. However, for the multi-zone climate control case, proper controller tuning is challenging, as externally, e.g., passenger-triggered changes in compressor setting and thus mass flow lead to degraded control performance. This paper presents a tuning method to automatically determine robust MPC parameters, as a function of the blower mass flow. Constrained contextual Bayesian optimization (BO) is used to derive policies minimizing a high-level cost function subject to constraints in a defined scenario. The proposed method leverages random disturbances and model-plant mismatch within the training episodes to generate controller parameters achieving robust disturbance rejection. The method contains a postprocessing step to achieve smooth policies that can be utilized in real-world applications. First, simulation results show that the mass flow-dependent policy outperforms a constant parametrization, %
       while achieving the desired closed-loop %
       behavior. Second, %
       the robust tuning method greatly reduces
       worst-case overshoot %
       and produces consistent closed-loop behavior under varying operating conditions. %
\end{abstract}

\begin{textblock*}{\textwidth}(0mm,-130mm)
	\small\textcopyright 2023 IEEE.  Personal use of this material is permitted.  Permission from IEEE must be obtained for all other uses, in any current or future media, including reprinting/republishing this material for advertising or promotional purposes, creating new collective works, for resale or redistribution to servers or lists, or reuse of any copyrighted component of this work in other works. 
\end{textblock*}

\section{Introduction}

Climate control for vehicles is important for reasons of comfort and safety. Thermally uncomfortable environments may lead to a decrease in driver's attention and a resulting increase in reaction times  \cite{Daly.2006}.  As a result, customers of modern vehicles, especially in higher-priced segments, expect a climate-control system capable of achieving high thermal comfort under varying operating conditions. 

Traditionally, model-free approaches such as proportional-integral-derivative (PID) control are used for this purpose. However, they require large design and application engineering efforts for manual tuning, and closed-loop performance may be unsatisfactory. This is caused in part by the large number of degrees of freedom of the resulting control architecture, introduced by necessary decoupling controllers. It has been shown (e.g., \cite{Wang.2018,Cvok.2021b, Reuscher.2021}) that model predictive controllers (MPC) have the potential to improve control performance and reduce the number of degrees of freedom especially for multi-zone temperature control. 

However, the correct tuning of the %
MPC hyper-parameters is %
critical for closed-loop control performance. In \cite{Reuscher.2021}, the authors presented an MPC structure for controlling the cabin temperature based on a user-selected mass flow. This control mode %
is challenging because the selected mass flow alters the system dynamics \cite{Poovendran.2020}. As a result, MPC parameters need to be adapted accordingly. %
Furthermore, unknown disturbances such %
solar irradiation or ambient temperature need to be robustly rejected by the MPC. %
However, their influence is hard to model accurately \cite{Zhang.2016b}. Necessary model order reductions for control can increase the model error further \cite{Klemm.2018} %
resulting in model-plant mismatch %
especially for lower-order control-oriented models. %
Due to the wide range %
in these disturbances, tuning a fixed parameter set by hand achieves satisfactory results only in selected operation scenarios.

Therefore, %
we propose to formulate the tuning problem as a constrained contextual optimization problem and approximately solve it with Bayesian optimization (BO). BO has been shown to be more sample-efficient than other black-box optimizers on various tuning problems in control \cite{Stenger.Bench}.

In literature, BO was used to tune the internal model of an MPC (e.g., \cite{Piga.2019}) and other hyper-parameters such as objective function weights parametrization and prediction horizon (e.g., \cite{DavidStenger.2020, Andersson.2016}) and Multi-objective BO was used to address conflicting objectives (\cite{Stenger.MO, MAKRYGIORGOS2022107770}). Robustly constrained optimization w.r.t. different model-plant mismatches has been considered for example in \cite{DavidStenger.2020}.

Contextual BO refers to the tuning of parameters as a function of operating conditions, in this case, the user-defined mass flow. Contextual BO for MPCs has seen one experimental application by \cite{Frohlich.06.10.2021}. However, constraints were not considered. Literature on constrained contextual BO for other controller structures used either a combination with safe BO \cite{Fiducioso.6282019} or approaches based on trust regions \cite{JinkyooPark.2020}. 

We propose to combine BO with Constrained Max-value Entropy Search (CMES) \cite{Perrone.15.10.2019} with contextual BO and therefore name this method C-CMES. %
A probabilistic constraint formulation ensures robustness against randomly generated disturbance trajectories and unknown model-plant mismatch. 
A post-processing step addresses smoothness requirements for the policy. 
The method, applied in simulation, highlights the practical applicability of automatic tuning based on BO for the domain of vehicle cabin climate control. %

This paper is structured as follows: In Section \ref{sec:MPC_tuningProb} the cabin temperature control tuning problem is described from an application point of view. Cost functions are defined and properties given. The optimization method proposed (C-CMES) is described in Section \ref{sec:BO_details}. Simulation results regarding performance, smoothness, and robustness are given in Section \ref{sec:results}. A discussion of results is given in Section \ref{sec:Conclusion}.

\newpage

\section{Problem Statement} \label{sec:MPC_tuningProb}

\subsection{MPC for Cabin Temperature Control}
\begin{figure}[b]
    \centering
    \includegraphics[trim=12 10 00 7,clip,scale = 1.2]{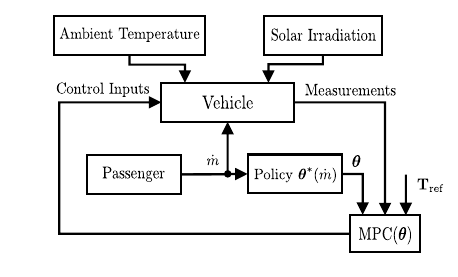}
    \caption{Signal flow diagram of the control structure presented in \cite{Reuscher.2021} augmented with mass flow-dependent policy. Blower mass flow as passenger-defined exogenous input for this contribution.}
    \label{fig:control_structure}
\end{figure}

The vehicle cabin temperature control system considered in this work is shown in Fig.~\ref{fig:control_structure}. The goal of the control loop is to track temperature references $T_{\mathrm{ref}}$ in three zones of the vehicle (zone 1: driver zone, zone 2: passenger zone, zone 3: font zone) and to effectively reject disturbances. This corresponds to a usual three-zone climatization setup \cite{Daly.2006}. For this purpose, a model predictive controller is used to modulate the temperature of inflowing air. The MPC's parameters $\params$ are chosen according to a policy (cf. \eqref{eq:optProblem}) based on the passenger-defined mass flow $\dot{m}$. The model underlying the control system was presented in \cite{Poovendran.2020} and is of the form:
\begin{equation}
    \begin{pmatrix}
    \T_{\mathrm{air}} \\ \T_{\mathrm{solids}}
    \end{pmatrix}_{p+1} = \g\left(\T_{\mathrm{air},p}, \T_{\mathrm{solids},p}, \dot m_p, \T_{\mathrm{mix},p} \right) \ ,
\end{equation}
with air temperature $\T_{\mathrm{air},p}$ at time step $p$, mean solid component temperature $\T_{\mathrm{solids},p}$, mass flow $\dot m_p$ and mixing temperature entering the climatization system $\T_{\mathrm{mix},p}$, each as a vector of three zonal values. The map $\g: \mathbb{R}^{12} \mapsto \mathbb{R}^6$ is strongly nonlinear in mass flow and slightly nonlinear in air temperature \cite{Poovendran.2020}. For the manual control mode of the cabin temperature, which the passenger can choose, the mixing temperature is used, while the mass flow level remains controlled by the vehicle passengers. For this control setup, the mass flow is thus an exogenous input. This baseline control system was initially presented in \cite{Reuscher.2020} with extensions in \cite{Reuscher.2021}. As the open loop is multiple-input multiple-output (MIMO) with strongly coupled states, but mostly linear for constant mass flows, an MPC is used with a continuously (i.e. in each time step) linearized system model. The cost function of this MPC is:
\begin{equation}
J_{\mathrm{MPC}} = \sum_{p=0}^{N_2-1}\norm{\T_p - \T_{\mathrm{ref},p}}{\mathbf{q}} + \norm{\Delta \T_{\mathrm{mix},p}}{\boldsymbol\lambda_p}  \ , \label{eq:MPC_cost_fun}
\end{equation}
with the MPC prediction horizon $N_2$, the cabin temperature reference $\T_{\mathrm{ref}}$, the change in mixing temperature between steps $\Delta \T_{\mathrm{mix},p} =\T_{\mathrm{mix},p} - \T_{\mathrm{mix},p-1} $, and weighting vector $\boldsymbol\lambda_p \in \mathbb{R}^{3}$. To achieve offset-free tracking, an extended Kalman filter is used to estimate input disturbances for all air temperature states. This filter is tuned by hand to find a compromise between convergence time and noise reduction. Details on observer tuning, the controller and more motivation can be found in \cite{Reuscher.2020, Reuscher.2021}.

\subsection{MPC Tuning Problem}
This section outlines the specific tuning problem addressed in this paper. However, the presented method treats the tuning problem as a black box. As a result, it can directly be applied to different degrees of freedom, training episodes, and high-level closed-loop control objectives. 
\paragraph{Degrees of Freedom} The MPC's degrees of freedom are here reduced to the matrix $\boldsymbol\lambda$, as horizons are predetermined by application needs and computation time limitations. Initial experiments have shown that $\lam_0$ should be chosen differently from the remaining entries to allow good tracking and disturbance rejection behavior (cf. \eqref{eq:lambda_matrix}). For simplicity, we set the tuning for the third zone to be constant. As the priority for both front zones is assumed to be identical, no separate tuning parameters for these zones were chosen. Therefore, the matrix $\lam \in \mathbb{R}^{3 \times N_2}$ for tuning is as follows:
\begin{equation} \label{eq:lambda_matrix}
 \lam = \begin{pmatrix}
     \lam_{p=0} & ... & \lam_{p=N_2-1}
 \end{pmatrix} = 
 \begin{pmatrix} 
     \lambda_{0} & \lambda & \dots & \lambda \\  
     \lambda_{0} & \lambda & \dots & \lambda \\
     1 & 1 & \dots & 1
\end{pmatrix} \ .
\end{equation}
We expect a similar influence of parameter changes in the log scale. This lines up with manual tuning experience and has been shown, e.g., by the authors in \cite{DavidStenger.2020}, Fig. 4.
We thus choose the tuning parameter vector to be $\params = \begin{pmatrix} \log(\lambda) & \log(\lambda_{0}) \end{pmatrix}$.

\paragraph{Simulative Training Episode} \label{sec:Epsisode}

A simulation environment is used to evaluate the MPC performance.   %
The optimizer repeatedly queries a simulative training episode with different values of the tuning parameters. The closed-loop behavior is then analyzed and fed back to the optimizer. This way, MPC parameters that achieve the desired closed-loop behavior consistently are searched for. 
\begin{figure}[bh]
    \centering
    \includegraphics{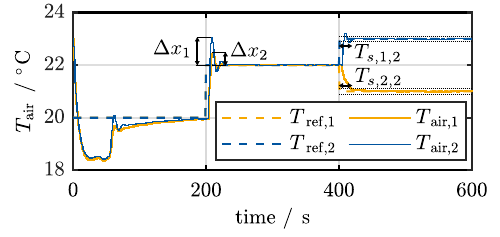}
    \caption{Training episode including characteristics in time domain}
    \label{fig:step_example}
\end{figure}

Fig. \ref{fig:step_example} depicts the simulative training episode used in this work. After an initial convergence of the disturbance observer, the reference changes in both front zones. Afterwards, a reference change leads to deviating zone temperatures. One simulative episode takes about \SI{30}{\second}.  

The simulation environment is designed such that it has similar properties as a real-world experiment. This way, parameters tuned in simulation may be applicable to the real system. Additionally, we can evaluate whether the tuning method can handle the challenges of real-world experiments such as noise, disturbances, and varying mass flows.

Therefore, the disturbances (environmental temperature and solar irradiation) are randomly sampled from a catalog of recorded trajectories for each objective function query. A random model plant mismatch is drawn for each function evaluation. All key parameters of the simulation model are uniformly sampled within $\pm$\SI{30}{\percent} of their nominal values. A constant mass flow is randomly chosen for each episode. In a commercial setting, this random choice may be made by test drivers before release or by customers after release.

\paragraph{Desired Closed-Loop Behaviour}

For vehicle cabin interior temperature controllers, first-order behaviour of the closed-loop control system is desirable (cf. Fig \ref{Fig:Case7_validation}, right). Independent of mass flow, disturbances, model-plant mismatch, step height or deviation between individual zones' setpoints, the reference shall be reached   %
\begin{enumerate}
    \item  with short settling time $T_s$,
    \item and limited overshoot $\Delta x$.
\end{enumerate}
Both characteristics are shown in Fig. \ref{fig:step_example}. %
Settling time and overshoot are defined for both step responses and both zones. To reduce both characteristics to a scalar value, the following definition is chosen:
\begin{equation}
    \Delta x = \max\left(\Delta x_1 , \Delta x_2\right) \qquad T_{\mathrm{s}} = \sum_{i,j} T_{\mathrm{s},i,j} \label{eq:xTs_definition}
\end{equation}
with overshoot $\Delta x_i$ for zone $i$  and the settling time $T_{\mathrm{s},i,j}$ for step $j$ and zone $i$ as a function of blower mass flow $\dot{m}$ and MPC parameters $\params$.

\subsection{Mathematical Optimization Problem Statement} \label{sec:mathProb}
To achieve the desired closed-loop behaviour, the tuning problem is formulated as a robustly constrained, contextual, and stochastic optimization problem: %
\begin{equation}
	\begin{aligned} \label{eq:optProblem} 
		\params^*(s) =  \argmin{\params \in \mathbb{R}^2} \qquad &  \mathbb{E} \left[ J(\params , s) \right] \\
		\mathrm{s.t. }\qquad \qquad \qquad \qquad  %
		& P  \left(g(\params , s) \le g_\mathrm{max} \right) \ge \delta \\
		& \mathbf{s}_{\mathrm{min}} \le  \mathbf{s} \le  \mathbf{s}_{\mathrm{max}}\\	
		& \params_{\mathrm{min}} \le  \params \le  \params_{\mathrm{max}} \ ,
	\end{aligned} %
\end{equation}
with maximum constraint $g_\mathrm{max}$ and constraint satisfaction probability $\delta$, context $s = \dot m$, its lower bound $\mathbf{s}_{\mathrm{min}}$, its upper bound $\mathbf{s}_{\mathrm{max}}$, optimal parameter vector $\params^*(s)$, parameter vector lower bound $\params_{\mathrm{min}}$, upper bound $\params_{\mathrm{max}}$. The constraints for $\mathbf{s}$ can be chosen based on physical relations and for $\params$ based on experience. The operators $\mathbb{E}$ and $P$ describe expected value and probability w.r.t. the randomly drawn disturbances and model-plant mismatch. %
Cost function $J(\params , s)$ and constraint $g(\params , s)$ are formulated as 
\begin{equation}
    J(\params , s) = T_{\mathrm{s}} \\ \qquad g(\params , s) = \Delta x \ . \label{eq:costfun_and_constraint}
\end{equation}
Note that $J(\params , s)$ relates to the hyperparameter optimization and is not the cost function of the MPC algorithm $J_\mathrm{MPC}$ in \eqref{eq:MPC_cost_fun}.
The key properties and design choices of \eqref{eq:optProblem}  %
are:

\paragraph*{Constrained vs. Multi-Objective Optimization}

Reducing settling time and overshoot (cf. \eqref{eq:xTs_definition}) are conflicting objectives. 
One way to address this conflict is to use single-objective unconstrained optimization with a weighted cost function. 
However, this generates an additional degree of freedom for the weighting between both terms. In practice, this results in tedious tuning by hand. As an alternative, Pareto optimization \cite{Stenger.MO} can be used to produce a set of compromises between the two objectives. However, this is expected to require more objective function evaluations.

Instead, we propose to separate the terms and use only settling time for the cost function and the overshoot as a constraint value. This formulation yields the desired first-order behavior, as the settling time cost achieves fast approaches to the reference, while the overshoot constraint effectively cancels out any oscillations or higher-order behavior.
Due to the probabilistic simulation, a robust constraint formulation 
\begin{equation} \label{eq:robustConstraint}
    P  \left(\Delta x(\params , \mdot) \le \Delta x_\mathrm{max} \right) \ge \delta
\end{equation}
is chosen in \eqref{eq:optProblem}. By choosing $\delta \stackrel{!}{=} 1$, the constraint can be guaranteed. This in practice leads to either infeasible behavior or very conservative tuning. Here, we chose $\delta = 0.5$ and $\delta = 0.93$ corresponding to $0~\sigma$ and $1.5~\sigma$ as non-robust and robust cases (cf. Fig.~\ref{Fig:Case7_PolicyComp}). %

\paragraph*{Contextual Optimization} The MPC's behavior heavily depends on the passenger-selected mass flow. Therefore, we search for optimal MPC parameters as a function of the mass flow, i.e., a policy $\params^*(\mdot): \mathbb{R} \mapsto \mathbb{R}^2$
instead of one optimal parameterization for all mass flows. This can be termed gain scheduling in classical control engineering terms.
Additionally, we require $\params^*(\mdot)$ to be smooth. Sudden parameter changes with changing mass flow may lead to undesired closed-loop switching behavior. Therefore, the second derivative of the solution $\frac{\mathrm{d}^2\params^*}{\mathrm{d}\mdot^2}$ should be small.

\paragraph*{Noisy Black-Box Optimization}
Information about the functions $J(\params , s)$ and $g(\params , s)$ can only be obtained by querying the simulation model with varying MPC parameters. Furthermore, evaluating the simulative episode twice with identical MPC parameters and context produces varying results, due to the randomly generated disturbances. The problem is thus a noisy black-box optimization problem. Critically, no analytical gradients can be given for the optimization. These properties restrict the class of usable optimization algorithms considerably. 

\paragraph*{Expensive Function Evaluations}

Each objective function evaluation is expensive as it takes around thirty seconds. Therefore, the optimization algorithm needs to be sample-efficient, i.e, it needs to be able to find good solutions with as little objective function evaluations as possible.

\paragraph*{Simple Regret and Safety} Here, we focus on the performance of the final policy $\params^*(\mdot)$ after $K$ iterations. %
Large cost function values during the optimization are not critical. Additionally, there are no immediate functional safety issues for %
exceeded constraint values during optimization. This is equivalent of minimizing simple regret instead of cumulative regret. Additionally, safe sampling is not required, although it can be achieved with BO \cite{Fiducioso.6282019}.

\newpage

\section{BO with C-CMES: Contextual Contrained Max-Value Entropy Search}
\label{sec:BO_details}

\subsection{Optimization}

BO has been shown to be more sample-efficient than other black-box optimizers %
\cite{Stenger.Bench}. Additionally, it naturally addresses noisy optimization problems. For a detailed introduction to BO, the reader is referred to \cite{garnett_bayesoptbook_2022} and \cite{Shahriari.2016}. Here, we extend constained max-value entropy search (CMES)\cite{Perrone.15.10.2019} to the contextual case to account for all challenges posed by \eqref{eq:optProblem}. This section emphasises the main differences to unconstrained single-objective BO. %
\begin{algorithm}[ht] 
	1: Generate an initial data set  $\mathcal{D}_1 = \left(\Theta_{1}, \objSet_{1}, \cnstrntSet_{1}, \contextSet_{1} \right)$ \\ [3pt] 
	2: \textbf{for} k = 1; 2; . . . ; K \textbf{do} \\[3pt]
	3: \quad  Learn probabilistic surrogate models $ \left(\mathcal{GP}_k^{J}, \mathcal{GP}_k^{G} \right)$ using all past evaluations 
	$\mathcal{D}_k = \left( \Theta_{k}, \objSet_{k}, \cnstrntSet_{k}, \contextSet_{k} \right)$ \\[3pt]
	4: \quad $\context_{k+1} \leftarrow \mathrm{receiveContext}()$ \\[3pt]
	5: \quad Select $\boldsymbol{\theta}^\prime_{k+1}$ by optimizing the acquisition function for context $\context_{k+1}$ : \\ 
	\begin{equation}
		\boldsymbol{\theta}^\prime_{k+1} = \argmin{\params_{\mathrm{min}} \le \boldsymbol{\theta} \le \params_{\mathrm{max}}} \ \mathrm{CMES}(\boldsymbol{\theta}, \context_{k+1}, \Theta_{k}, \mathcal{GP}_k^{G}, \mathcal{GP}_k^{J})
	\end{equation}	
	6: \quad Query expensive-to-evaluate simulation with $\boldsymbol{\theta}^\prime_{k+1}$ to obtain $j^\prime_{k+1}$ and $g^\prime_{k+1}$ for context $\context_{k+1}$ \\[3pt]
	7: \quad Augment data set with new evaluations to obtain \hbox{$\mathcal{D}_{k+1} = ( \paramSet_{k+1},  \objSet_{k+ 1},\cnstrntSet_{k+ 1}, \contextSet_{k+1}) $}\\[3pt] 
	8: \textbf{end for} \\[3pt]
	9: Obtain final smooth policy $\params^*(s)$ (Sec. \ref{sec:postproc})
	\caption{Constrained Contextual BO. \label{Algo:BayesOpt}}
\end{algorithm}

First, in Step 1 of Algo. \ref{Algo:BayesOpt}, an initial data set $\mathcal{D}_1$ consisting of evaluated parameters $\Theta_{1}$ for contexts $\contextSet_{1}$ and obtained objective function $\objSet_{1}$ and constraint $\cnstrntSet_{1}$ values is generated to obtain initial information about the latent objective function $J(\params,\context)$ and constraints $g(\params,\context)$. Latent means that only noisy samples of both functions are available (cf. Sec. \ref{sec:mathProb}). 

The main optimization loop spans Steps 2 - 8. In each iteration Gaussian Process (GP) surrogate models of objective function and constraint function are generated (Step 3):
\begin{equation}
	\begin{aligned}
		J(\boldsymbol{\theta},\context) &\approx \tilde{J}_k(\boldsymbol{\theta},\context \mid \mathcal{D}_k) \sim \mathcal{N} \left(\mu_{\mathrm{J},k}(\boldsymbol{\theta},\context),\sigma^2_{\mathrm{J},k}(\boldsymbol{\theta},\context)\right), \\
		g(\boldsymbol{\theta},\context) &\approx \tilde{g}_{k}(\boldsymbol{\theta},\context \mid \mathcal{D}_k) \sim \mathcal{N} \left(\mu_{\mathrm{g},k}(\boldsymbol{\theta},\context),\sigma^2_{\mathrm{g},k}(\boldsymbol{\theta},\context)\right).	
	\end{aligned}
\end{equation}
We follow \cite{Krause.2011} in modelling the true unknown objective function $J(\boldsymbol{\theta},\context)$ as a GP. The model of iteration $k$, $\mathcal{GP}_k^{J}$, yields normally distributed predictions $\tilde{J}_k(\boldsymbol{\theta},z \mid \mathcal{D}_k)$ as a function of parameters and context. Predictive mean and predictive standard deviation are denoted as $\mu$ and $\sigma$, respectively. Predictive uncertainty increases in areas of the parameter-context space, where the MPC performance has not been evaluated yet. For the GP models (cf. \cite{CarlEdwardRasmussenandChristopherK.I.Williams.2006}), a constant priori mean, homeoscedastic Gaussian likelihood, and an anisotropic squared exponential is used. Smooth box hyper-priors are placed on the kernel length scales in order to constrain them to sensible orders of magnitude. The GP models hyperparameters are updated at each iteration using maximum a-posteriori estimation. For the constrained function $g(\boldsymbol{\theta},\context)$ an independent GP model with identical settings is constructed. 

The GP models are used to determine useful parameters $\params'_{k+1}$ to be evaluated next. This decision also depends on the context of the next episode. As described in Sec. \ref{sec:MPC_tuningProb}, the context cannot be chosen by the optimizer and is instead received from the simulation environment (Step 4). When choosing the next sample point, a so-called acquisition function is used to balance between exploration and exploitation (Step  5). Here, we use CMES \cite{Perrone.15.10.2019}, an information theoretic acquisition function, based on max-value entropy search (\cite{Wang.06.03.2017}). In contrast to improved-based acquisition functions, CMES does not require a single \textit{current best solution}.
Additionally, in contrast to UCB that was used in \cite{Frohlich.06.10.2021} and \cite{JinkyooPark.2020}, CMES naturally handles constraints and therefore does not require trust regions or SafeBO. 
In general, the acquisition function is multimodal. Therefore, it is optimized using a combination of random search and gradient-based optimization.

In Step 6, the expensive-to-evaluate simulative episode is queried with the suggested controller parameters in order to obtain the resulting system responses. After the data set is augmented (Step 7), the next iteration is started. The optimization is terminated after a given budget of $K$ objective function evaluations is exhausted.   

\subsection{Post Processing for Smoothness} \label{sec:postproc}

After the main optimization loop has terminated, final GP models $\mathcal{GP}_K^{J}, \mathcal{GP}_K^{G}$ are available. Although they ideally have been refined in promising parameter regions, a final smooth policy $\params^*(s)$ still needs to be obtained. 

To achieve that, we discretize the context space $\contextSet_{\mathrm{disc}} = \{s_1,...,s_N\}$. The corresponding optimal parameter values are denoted as $\boldsymbol{\theta}_{1}^* , \dots, \boldsymbol{\theta}_{N}^*$. To obtain them the expected objective function value is minimized, while the probabilistic constraints are required to hold:
\begin{equation}
	\begin{aligned} \label{eq:PostProcessing} 
		&  \min_{\boldsymbol{\theta}_{1}, \dots, \boldsymbol{\theta}_{N}  \in \mathbb{R}^d} \quad    \sum_{n = 1}^{N} \mu_{\mathrm{J},K}(	\boldsymbol{\theta}_{n},s_n) \ + ... \\
        & \qquad \qquad \qquad \qquad ... \gamma \sum_{n=1}^{N-2} \left( \params_{n+2} - 2 \params_{n+1} + \params_{n} \right) \\ %
		\mathrm{s.t. }\quad  & \params_{\mathrm{min}} \le \boldsymbol{\theta} \le \params_{\mathrm{max}}\\
		& \psi \left(\frac{\mu_{\mathrm{g},K}(\boldsymbol{\theta}_{n},s_n) - g_\mathrm{max}}{\sigma_{\mathrm{g},K}(\boldsymbol{\theta}_{n},s_n)} \right) > \delta, \, n \in \{1, \dots, N\}, 
	\end{aligned}
\end{equation}
As an additional term, the second-order finite difference of the discretized optimal policy 
addresses the smoothness requirements. The weighting parameter $\gamma$ needs to be adjusted manually, to get the desired smoothness. In practice, the discretized policy can be interpolated in order to extract parameters for each context.
Eq. \eqref{eq:PostProcessing} is solved using gradient-based optimization. The optimal feasible solution for each of the discretized contexts is used as an initial guess.

\section{Simulation Results}
\label{sec:results}
\subsection{Policy Smoothing}
\begin{figure} [ht]
	\begin{minipage}[t]{.48\textwidth}
		\centering
		\includegraphics[width=\linewidth]{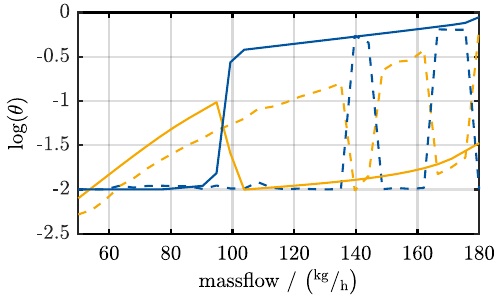}
	\end{minipage}
	\hfill
	\begin{minipage}[t]{.48\textwidth}
		\centering
		\includegraphics[width=\linewidth]{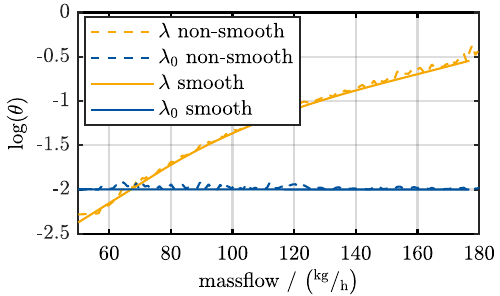}
	\end{minipage}
	\caption[Visualization of the optimized MPC parameters for vehicle cabin climate control]{Optimized context-/mass flow-dependent MPC parameterizations. \textbf{Top}: Robust policy ($\delta = 0.93$). \textbf{Bottom}: Non-robust policy($\delta = 0.5$).}
	\label{Fig:Case7_PolicyComp}
\end{figure}

The policy retrieved after \SI{500}{} black-box evaluations is shown in Fig. \ref{Fig:Case7_PolicyComp}. Both smooth and non-smooth policies, e.g., MPC parameters as a function of mass flow, are shown for the robust case in the top plot and the non-robust case (cf. \eqref{eq:robustConstraint}) in the bottom plot. For the non-robust case, the smoothing postprocessing leads to a mere smoothing of the solution. In the robust case, the smoothing fixes the solution to one of the optimal solutions between which the non-smooth solution appears to be switching. Up to about \SI{90}{\kilogram \per \hour}, the behavior is similar to the non-robust case. Afterwards, $\lambda$ drops, while $\lambda_0$ increases sharply. The non-smooth policy switches multiple times between these two solutions. Without smoothing, this behavior would not be acceptable for a closed-loop operation of the controller.

\subsection{Constant vs. Contextual Policy}

\begin{figure*} [t]
	\begin{minipage}[t]{.48\textwidth}
		\centering
		\includegraphics[width=\linewidth]{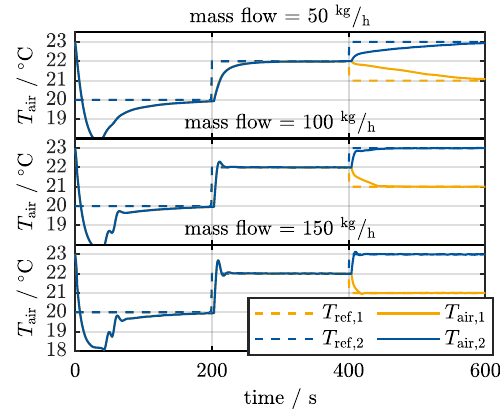}
	\end{minipage}
	\hfill
	\begin{minipage}[t]{.50\textwidth}
		\centering
		\includegraphics[width=\linewidth]{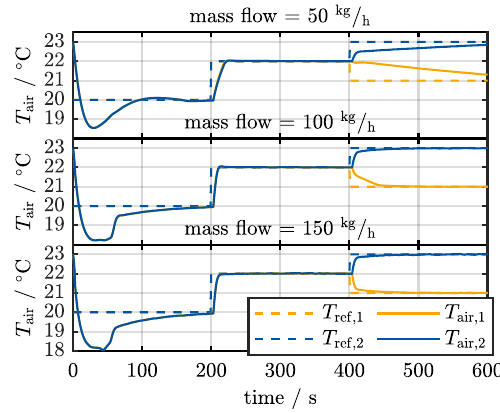}
	\end{minipage}
			\caption[Comparison of constant and contextual MPC parametrization for vehicle cabin climate control]{Closed-loop time domain plots for the training episode with different mass flows and constant model-plant mismatch and disturbance. \textbf{Left}: Constant MPC parametrization. Overshoot is visible after reference steps.  \textbf{Right}: Smooth and non-robust contextual MPC parametrization. Only minor overshoot after reference steps.}
	\label{Fig:Case7_validation}
\end{figure*}

The advantage of contextual over constant tuning is shown in Fig. \ref{Fig:Case7_validation}. Using constant tuning, only at $\dot m = \SI{50}{\kilogram \per \hour}$ the desired closed-loop behavior was achieved. The same tuning leads to \SI{0.3}{\kelvin} overshoot at $\dot m = \SI{100}{\kilogram \per \hour}$ and unwanted oscillations, as well as \SI{0.7}{\kelvin} overshoot at $\dot m = \SI{150}{\kilogram \per \hour}$. In contrast, contextual tuning achieves satisfactory results for all mass flows. %
No overshoot and first-order behavior can be seen for all mass flows. Thus, contextual tuning is highly important to cabin temperature control.

\subsection{Robustness}
\begin{figure} [ht]
	\begin{minipage}[t]{.24\textwidth}
		\centering
		\includegraphics[width=\linewidth]{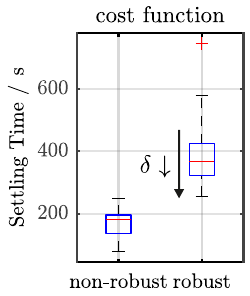}
	\end{minipage}
	\hfill
	\begin{minipage}[t]{.24\textwidth}
		\centering
		\includegraphics[width=\linewidth]{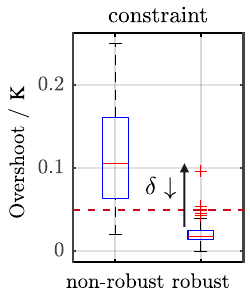}
	\end{minipage}
	\caption[Comparison of the cumulative overhead of different MOBO variants]{Distribution of objective function and constraint values obtained from 50 validation samples. Dotted red line: allowed overshoot of \SI{0.05}{\kelvin}}
	\label{Fig:Case7_boxplots}
\end{figure}

To validate the robust behavior of the closed control loop, %
50 episodes with model plant mismatch and disturbance sampling (cf. Sec. \ref{sec:Epsisode}) are simulated for the two smooth policies shown in Fig. \ref{Fig:Case7_PolicyComp}. The individual model plant mismatch combinations were different to the ones drawn in policy generation. The results of these simulations are shown in Fig. \ref{Fig:Case7_boxplots}. The distribution of cost function (i.e. settling time) values is shown in the top plot. The robust set shows overall a worse performance regarding cost function. Both mean and variance are higher. For the constraint shown in the bottom plot, mean and especially variance is strongly reduced. \SI{94}{\percent} of samples were below the threshold of \SI{0.05}{\kelvin}, which matches the desired value of $\delta = \SI{93.7}{\percent}$. The increased robustness regarding constraint compliance of the method is thereby validated. The case shown here puts a strong focus on robustness. The effect of decreasing delta and thus relaxing this focus is also shown in Fig. \ref{Fig:Case7_boxplots}. The method allows an application-driven trade-off between the characteristics.

\section{Conclusion}
\label{sec:Conclusion}

This paper proposed a Bayesian optimization methodology that produces controller parameters, i.e., a policy, as a smooth function of the context. Furthermore, the policy is robust w.r.t. randomly drawn disturbance trajectories minimizing a high-level objective while fulfilling constraints. 

The method is used to optimize the parameters of an MPC for vehicle climate control. Three main results are obtained from Monte Carlo validation simulations. First, a distinct post-processing step allows obtaining a smooth policy from the GP models fitted during optimization. Second, contextual optimization is elemental for operating an MPC under different conditions, in this case, different blower mass flows. Third, a robust formulation of the constraints enables BO to generate a policy that is robust to different ambient temperature and solar irradiation trajectories.

The results encourage an experimental application of the presented method on a physical vehicle in future work. The simulative training episode used here has the same structural properties as a real-world experiment, therefore easy transferabilty is expected. Additionally, other BO versions have been shown to be applicable in experiments in other applications, e.g. \cite{Frohlich.06.10.2021}.

\bibliography{IEEEabrv, ifacconf}

\begin{thebibliography}{10}
\providecommand{\url}[1]{#1}
\csname url@rmstyle\endcsname
\providecommand{\newblock}{\relax}
\providecommand{\bibinfo}[2]{#2}
\providecommand\BIBentrySTDinterwordspacing{\spaceskip=0pt\relax}
\providecommand\BIBentryALTinterwordstretchfactor{4}
\providecommand\BIBentryALTinterwordspacing{\spaceskip=\fontdimen2\font plus
\BIBentryALTinterwordstretchfactor\fontdimen3\font minus
  \fontdimen4\font\relax}
\providecommand\BIBforeignlanguage[2]{{%
\expandafter\ifx\csname l@#1\endcsname\relax
\typeout{** WARNING: IEEEtran.bst: No hyphenation pattern has been}%
\typeout{** loaded for the language `#1'. Using the pattern for}%
\typeout{** the default language instead.}%
\else
\language=\csname l@#1\endcsname
\fi
#2}}

\bibitem{Daly.2006}
S.~Daly, \emph{Automotive air-conditioning and climate control systems},
  1st~ed.\hskip 1em plus 0.5em minus 0.4em\relax Amsterdam:
  Butterworth-Heinemann, 2006.

\bibitem{Wang.2018}
H.~Wang, I.~Kolmanovsky, M.~R. Amini, and J.~Sun, ``Model predictive climate
  control of connected and automated vehicles for improved energy efficiency,''
  in \emph{2018 Annual American Control Conference (ACC)}.\hskip 1em plus 0.5em
  minus 0.4em\relax Piscataway, NJ: IEEE, 2018.

\bibitem{Cvok.2021b}
I.~Cvok, B.~{\v{S}}kugor, and J.~Deur, ``Control trajectory optimisation and
  optimal control of an electric vehicle hvac system for favourable efficiency
  and thermal comfort,'' \emph{Optimization and Engineering}, vol.~22, no.~1,
  pp. 83--102, 2021.

\bibitem{Reuscher.2021}
T.~Reuscher, K.~Poovendran, and D.~Abel, ``Model predictive zonal temperature
  control of a vehicle cabin,'' in \emph{2021 IEEE/ASME International
  Conference on Advanced Intelligent Mechatronics (AIM)}, 2021, pp. 37--43.

\bibitem{Poovendran.2020}
K.~Poovendran, D.~Abel, T.~Reuscher, and V.~Govender, ``Vehicle cabin thermal
  multi-zone modelling for control,'' in \emph{2020 2nd International
  Conference on Control Systems, Mathematical Modeling, Automation and Energy
  Efficiency (SUMMA)}.\hskip 1em plus 0.5em minus 0.4em\relax IEEE, 11/11/2020
  - 11/13/2020, pp. 489--495.

\bibitem{Zhang.2016b}
Q.~Zhang, S.~E. Li, and K.~Deng, \emph{Automotive Air Conditioning}.\hskip 1em
  plus 0.5em minus 0.4em\relax Cham: {Springer International Publishing}, 2016.

\bibitem{Klemm.2018}
D.~Klemm, W.~Roessner, N.~Widdecke, and J.~Wiedemann, ``Reduced model of a
  vehicle cabin for transient thermal simulation,'' ser. SAE Technical Paper
  Series.\hskip 1em plus 0.5em minus 0.4em\relax {SAE International}, 2018.

\bibitem{Stenger.Bench}
\BIBentryALTinterwordspacing
D.~Stenger and D.~Abel, ``Benchmark of bayesian optimization and metaheuristics
  for control engineering tuning problems with crash constraints,'' 2022.
  [Online]. Available: \url{https://arxiv.org/abs/2211.02571}
\BIBentrySTDinterwordspacing

\bibitem{Piga.2019}
D.~Piga, M.~Forgione, S.~Formentin, and A.~Bemporad, ``Performance-oriented
  model learning for data-driven mpc design,'' \emph{IEEE Control Systems
  Letters}, vol.~3, no.~3, pp. 577--582, 2019.

\bibitem{DavidStenger.2020}
D.~Stenger, M.~Ay, and D.~Abel, ``Robust parametrization of a model predictive
  controller for a cnc machining center using bayesian optimization,''
  \emph{IFAC-PapersOnLine}, vol.~53, no.~2, pp. 10\,388--10\,394, 2020.

\bibitem{Andersson.2016}
O.~Andersson, M.~Wzorek, P.~Rudol, and P.~Doherty, ``Model-predictive control
  with stochastic collision avoidance using bayesian policy optimization,'' in
  \emph{2016 IEEE International Conference on Robotics and Automation
  (ICRA)}.\hskip 1em plus 0.5em minus 0.4em\relax IEEE, 2016, pp. 4597--4604.

\bibitem{Stenger.MO}
\BIBentryALTinterwordspacing
D.~Stenger, R.~Ritschel, F.~Krabbes, R.~Voßwinkel, and H.~Richter, ``What is
  the best way to optimally parameterize the mpc cost function for vehicle
  guidance?'' \emph{Mathematics}, vol.~11, no.~2, 2023. [Online]. Available:
  \url{https://www.mdpi.com/2227-7390/11/2/465}
\BIBentrySTDinterwordspacing

\bibitem{MAKRYGIORGOS2022107770}
G.~Makrygiorgos, A.~D. Bonzanini, V.~Miller, and A.~Mesbah,
  ``Performance-oriented model learning for control via multi-objective
  bayesian optimization,'' \emph{Computers \& Chemical Engineering}, vol. 162,
  p. 107770, 2022.

\bibitem{Frohlich.06.10.2021}
\BIBentryALTinterwordspacing
L.~P. Fr{\"o}hlich, C.~K{\"u}ttel, E.~Arcari, L.~Hewing, M.~N. Zeilinger, and
  A.~Carron, ``Model learning and contextual controller tuning for autonomous
  racing.'' [Online]. Available: \url{https://arxiv.org/pdf/2110.02710}
\BIBentrySTDinterwordspacing

\bibitem{Fiducioso.6282019}
\BIBentryALTinterwordspacing
M.~Fiducioso, S.~Curi, B.~Schumacher, M.~Gwerder, and A.~Krause, ``Safe
  contextual bayesian optimization for sustainable room temperature pid control
  tuning.'' [Online]. Available: \url{http://arxiv.org/pdf/1906.12086v1}
\BIBentrySTDinterwordspacing

\bibitem{JinkyooPark.2020}
\BIBentryALTinterwordspacing
{Jinkyoo Park}, ``Contextual bayesian optimization with trust region (cbotr)
  and its application to cooperative wind farm control in region 2,''
  \emph{Sustainable Energy Technologies and Assessments}, vol.~38, p. 100679,
  2020. [Online]. Available:
  \url{https://www.sciencedirect.com/science/article/pii/S2213138819303108}
\BIBentrySTDinterwordspacing

\bibitem{Perrone.15.10.2019}
\BIBentryALTinterwordspacing
V.~Perrone, I.~Shcherbatyi, R.~Jenatton, C.~Archambeau, and M.~Seeger,
  ``Constrained bayesian optimization with max-value entropy search.''
  [Online]. Available: \url{https://arxiv.org/pdf/1910.07003}
\BIBentrySTDinterwordspacing

\bibitem{Reuscher.2020}
T.~Reuscher, K.~Poovendran, and D.~Abel, ``Control oriented modeling and
  nonlinear model predictive temperature control of a vehicle cabin,'' in
  \emph{European Control Conference 2020}, A.~Pogromsky and Y.~Ebihara,
  Eds.\hskip 1em plus 0.5em minus 0.4em\relax Piscataway, NJ: IEEE, 2020, pp.
  124--129.

\bibitem{garnett_bayesoptbook_2022}
R.~Garnett, \emph{{Bayesian Optimization}}.\hskip 1em plus 0.5em minus
  0.4em\relax Cambridge University Press, 2022, in preparation.

\bibitem{Shahriari.2016}
B.~Shahriari, K.~Swersky, Z.~Wang, R.~P. Adams, and N.~de~Freitas, ``Taking the
  human out of the loop: A review of bayesian optimization,'' \emph{Proceedings
  of the IEEE}, vol. 104, no.~1, pp. 148--175, 2016.

\bibitem{Krause.2011}
\BIBentryALTinterwordspacing
A.~Krause and C.~S. Ong, ``Contextual gaussian process bandit optimization,''
  in \emph{Advances in Neural Information Processing Systems 24}, {J.
  Shawe-Taylor}, {R. S. Zemel}, {P. L. Bartlett}, {F. Pereira}, and {K. Q.
  Weinberger}, Eds.\hskip 1em plus 0.5em minus 0.4em\relax {Curran Associates,
  Inc}, 2011, pp. 2447--2455. [Online]. Available:
  \url{http://papers.nips.cc/paper/4487-contextual-gaussian-process-bandit-optimization.pdf}
\BIBentrySTDinterwordspacing

\bibitem{CarlEdwardRasmussenandChristopherK.I.Williams.2006}
\BIBentryALTinterwordspacing
C.~E. Rasmussen and C.~K.~I. Williams, \emph{Gaussian Processes for Machine
  Learning}, 2006. [Online]. Available:
  \url{http://www.gaussianprocess.org/gpml/}
\BIBentrySTDinterwordspacing

\bibitem{Wang.06.03.2017}
\BIBentryALTinterwordspacing
Z.~Wang and S.~Jegelka, ``Max-value entropy search for efficient bayesian
  optimization.'' [Online]. Available: \url{https://arxiv.org/pdf/1703.01968}
\BIBentrySTDinterwordspacing

\end{thebibliography}

\end{document}